\documentclass[pdflatex,sn-vancouver,Numbered]{sn-jnl}%
\usepackage{graphicx}%
\usepackage{multirow}%
\usepackage{amsmath,amssymb,amsfonts}%
\usepackage{amsthm}%
\usepackage{mathrsfs}%
\usepackage[title]{appendix}%
\usepackage{xcolor}%
\usepackage{textcomp}%
\usepackage{manyfoot}%
\usepackage{booktabs}%
\usepackage{algorithm}%
\usepackage{algorithmicx}%
\usepackage{algpseudocode}%
\usepackage{listings}%
\usepackage{enumerate}%
\usepackage{cases}%
\usepackage{xcolor}%
\usepackage{tikz}
\usetikzlibrary{positioning} 
\usetikzlibrary{arrows} 
\usepackage{algorithm}
\usepackage{algpseudocode}

\newcommand{\func}{\textrm}
\let\oldReturn\Return
\renewcommand{\Return}{\State\oldReturn}
\newcommand{\1}{\mathbf{1}}

\algrenewcommand\algorithmicrequire{\textbf{Input:}}
\algrenewcommand\algorithmicensure{\textbf{Output:}}

\newtheorem{theorem}{Theorem}%
\newtheorem{statement}[theorem]{Statement}%

\theoremstyle{definition}%

\newcommand{\norm}[1]{\left\lVert#1\right\rVert}%

\begin{document}

\title[Article Title]{A Quantum Constraint Generation Framework for Binary Linear Programs}

\author*[1]{\fnm{Andr\'as} \sur{Cz\'egel}}\email{czegel@inf.u-szeged.hu}

\author[1]{\fnm{Bogl\'arka} \sur{G.-T\'oth}}\email{boglarka@inf.szte.hu}

\affil*[1]{\orgdiv{Department of Computational Optimization}, \orgname{University of Szeged}, \orgaddress{\street{\'Arp\'ad t\'er 2.}, \city{Szeged}, \postcode{6720}, \country{Hungary}}}

\abstract{
We propose a new approach to utilize quantum computers for binary linear programming (BLP), which can be extended to general integer linear programs (ILP). Quantum optimization algorithms, hybrid or quantum-only, are currently general purpose, standalone solvers for ILP. However, to consider them practically useful, we expect them to overperform the current state of the art classical solvers. That expectation is unfair to quantum algorithms: in classical ILP solvers, after many decades of evolution, many different algorithms work together as a robust machine to get the best result. This is the approach we would like to follow now with our quantum 'solver' solutions. In this study we wrap any suitable quantum optimization algorithm into a quantum informed classical constraint generation framework. First we relax our problem by dropping all constraints and encode it into an Ising Hamiltonian for the quantum optimization subroutine. Then, by sampling from the solution state of the subroutine, we obtain information about constraint violations in the initial problem, from which we decide which coupling terms we need to introduce to the Hamiltonian. The coupling terms correspond to the constraints of the initial binary linear program. Then we optimize over the new Hamiltonian again, until we reach a feasible solution, or other stopping conditions hold. Since one can decide how many constraints they add to the Hamiltonian in a single step, our algorithm is at least as efficient as the (hybrid) quantum optimization algorithm it wraps. We support our claim with results on small scale minimum cost exact cover problem instances. 
}

\maketitle

\section{Introduction}

Optimization algorithms have been in the headlines of advances towards the utility of quantum computation for many years. Recently, a great state of the art collaborative article \cite{Abbas2024} of many leading researchers of variational quantum algorithms and Operations Research (OR) came out, summarizing where the field of Quantum Optimization is heading, what current advances and challenges are. 

One point stands out even from their work: current quantum algorithms are mostly standalone routines \cite{McArdle2019, Cubitt2023, Kadowaki1998}. Hybrid algorithms slightly differ as they are like most global optimization algorithms, use two algorithms: one hyper-parameterized, usually local, solver and an optimizer aiming to find optimal solutions for the original problem by guiding the local solver \cite{Peruzzo2014, Farhi2014, Hadfield2019, Bravyi2020}. Since finding new algorithms \cite{Gacon2024, Kyaw2023, Dalzell2023, Motta2019, Soley2021}, 
driving the quantum optimization community, algorithmic performance is still far from the expectations of OR practitioners. One might argue that current NISQ devices are not suitable for such tasks \cite{Saxena2024} and for certain problems never will be \cite{Stoudenmire2024}, or that these devices are not meant to solve ILP instances and not even suitable for the mathematical problem they pose. One could form the question what advances the field of quantum technology and what kind of new algorithms we need to catch up? We propose another question: what have we taken away from the long decades of evolution in ILP solving techniques to solve problems with the help of quantum devices? More importantly, what ideas are there to improve our existing quantum solving techniques?

We would take a step back from practical issues of hardware and noise and would like to take a look into this direction. Current state of the art (Mixed) ILP solvers consist of a large number of exact and heuristic algorithms for cleverly constructed subproblems. Their core usually is still some branching algorithm and a linear programming (LP) solver method, however they have evolved far from being one single algorithm. We don't compare current quantum algorithms to an implementation of the original Cutting Planes \cite{Gomory1958} idea, nor to a rudimentary realization of Branch and Bound \cite{Land1960}.
Of course we do not, because they had been introduced more than 60 years ago. And imagine running them on the hardware scientists had in the late 50's! We would like to see advances in solving our current problems, even though the state of of quantum optimization is not there yet. 

There are many exciting studies into this direction considering quantum speedup for tree search, like B\&B algorithms \cite{Chakrabarti2022, Montanaro2020}, applying Benders decomposition for MILPs to split the problem into binary programming and LP parts \cite{Chang2020, Zhao2022} and solving the binary on a quantum device. And there is also a primal-dual method based on Lagrangian relaxation and composition of original variables and variational quantum circuit parameters \cite{Le2023}.

Our study addresses these questions by putting together some puzzle pieces. We take existing quantum optimization algorithms, take the idea of constraint generation, and their 'common sides' as a quantum advice obtained from sampling. We describe the framework formally in Section \ref{sec:congen}, then show experimental results in Section \ref{sec:res}.  Discussion of the results are given in Section \ref{sec:discussion}, while Section \ref{sec:conclusion} concludes the paper.

\section{Constraint generation algorithm}
\label{sec:congen}

\subsection{Problem definition}

Our initial binary linear problem (BLP) is to find

\begin{equation}\label{orig}
    \min~ \left\{ c^Tx ~\bigm|~ Ax = b, x\in\left\{0,1\right\}^n \right\}, \tag{BLP}
\end{equation}
where $c\in\mathbb{R}^n, A\in\mathbb{R}^{m \times n}, b\in\mathbb{R}^m$. As BLP is NP-complete \cite{Karp1972}, there exists an efficient reduction from generic Integer Linear Programs (ILPs) to BLPs \cite{Lucas2014}. Thus, our algorithm is suitable for solving any ILP by reformulating it as a BLP. In the following sections we propose an algorithmic scheme to solve (BLP).

\subsection{Constructing the physical problem}
\label{sec:problemdef}

From the binary linear problem, we need to construct the problem that we would solve via a quantum subroutine. The conversion itself is an important ingredient of our results, however it is fairly straightforward. The methodology is that we first convert the problem into a quadratic, unconstrained problem (QUBO), on which we introduce how the relaxation of the original constraints can be done. Then, the relaxed unconstrained quadratic program is transformed into an Ising Hamiltonian. We chose this path because it allows us to perform constraint generation of the initial linear problem (BLP) directly to the Hamiltonian. That means the constraints of (BLP) appear as coupling terms in the physical system, which also implies that our initial physical problem is also trivial and that the constraint generation algorithm gradually adds complexity to the Hamiltonian until, in the worst case, it reaches its full complexity. That is essentially the straight BLP to Ising Hamiltonian conversion of the (BLP) without relaxations. 

\subsubsection{Relaxed Quadratic Program}

The next step is to transform (BLP) into a Quadratic Unconstrained Binary Optimization (QUBO) problem, i.e. to the form
\begin{equation}\label{quboform}
    \min~ \left\{ x^TQx + c^Tx + d ~\bigm|~ x\in\left\{0,1\right\}^n \right\},
\end{equation}
where $Q \in\mathbb{R}^{n \times n}, Q_{ij} = Q_{ji}$ and $d$ is a real constant. The term $c^Tx$ is usually encoded into $x^TQx$, but we would like to keep it this way as it will explicitly show added complicating terms in the Hamiltonian.

The transformation is simply to obtain a second order penalty term from the constraints of (BLP) 
\begin{align}
    Ax=b ~\longmapsto& ~\quad (Ax-b)^T(Ax-b) \\ 
    &= x^TA^TAx - 2b^TAx + b^Tb \\ 
    &= x^T(A^TA)x - (2b^TA)x + b^Tb
\end{align}
and adding them to the objective as a penalty, weighted by a big $M$, the problem becomes 
\begin{equation}
    \min~ \left\{x^T(MA^TA)x + (c^T - 2Mb^TA)x + Mb^Tb ~\bigm|~ x\in\left\{0,1\right\}^n\right\},
\end{equation}
that has the desired form.
The form has one caveat, that is, $M$ should be a sufficiently large constant. We describe how to calculate a suitable value for $M$ in Appendix \ref{app:M}.

Now, let us define $\hat{A}\in\mathbb{R}^{m \times n}$ as our \textit{penalty} matrix by selecting given rows from $A$ and $\hat{b} \in \mathbb{R}^m$ as the corresponding constant terms of the constraints. The matrix $\hat{A}$ contains our constraints that we add to the problem. 

Let us define our relaxed quadratic problem (RQP) as
\begin{equation}
    \min~ \left\{x^T(M\hat{A}^T\hat{A})x + (c^T - 2M\hat{b}^T\hat{A})x + M\hat{b}^T\hat{b}~\bigm|~ x\in\left\{0,1\right\}^n\right\}
    \tag{RQP}
\end{equation}

In the initial step, we define $\hat{A}^{(0)}$ as the zero matrix and $\hat{b}^{(0)}$ as the zero vector with the respective dimensions. Thus, (RQP) resolves to the trivial problem of
\begin{equation}
    \min \left\{c^Tx ~\bigm|~ x\in\left\{0,1\right\}^n\right\}.\label{eq:trivi}
\end{equation}
Transforming \eqref{eq:trivi} into the Ising Hamiltonian results in a trivial problem without coupling terms. However, it is an essential starting step for our algorithm, to compute violation scores and add the first constraints.

\subsubsection{Relaxed Problem Hamiltonian}
\label{sec:RPH}

The Hamiltonian we would like to perform a ground state search on is an Ising Hamiltonian, created directly from (RQP). The Ising Hamiltonian, for $\sigma \in \{1, -1\}^n$,
is given by
\begin{equation}
    H(\sigma) = - \sum_{ij}J_{ij}\sigma_i \sigma_j - \mu \sum_i h_i \sigma_i.
\end{equation}
Starting from (RQP), by converting the variables' domain from $\{1, 0\}$ to $\{1, -1\}$ and dropping the constant term that does not affect the the optimal solution, we obtain that the coefficients in matrix form are 
\begin{align}
    J &= -\frac{1}{4}M\hat{A}^T\hat{A}, \label{eq:J}\\[6pt]
    h &= c^T - 2 M\hat{b}^T \hat{A} + M\mathbf{1}^T (\hat{A}^T \hat{A}), \\[6pt]
    \mu &= -\frac{1}{2}, \label{eq:mu}
\end{align}
where $\mathbf{1}$ is the all-one vector of the respective dimension.
This way we arrive at the final form of the Relaxed Problem Hamiltonian (RPH) as
\begin{equation}
    \min \left\{H(\sigma) ~\bigm|~ \sigma \in \{1,-1\}^n \right\},
    \tag{RPH}
\end{equation}
i.e.\ looking for the lowest energy spin configuration of $H(\sigma)$. See Appendix \ref{app:QUBOtoHAM} for derivations to obtain \eqref{eq:J}-\eqref{eq:mu}. 

We would like to point out a few things here. First, our initial step has no coupling constraints, which means that no two or more qubit interactions are present in our Hamiltonian. Secondly, as we add constraints into our initial linear model, and so to (RQP), that is equivalent to adding coupling terms to our Hamiltonian while also modifying the strength of the magnetic field of single qubits. Since each constraint complicates the Hamiltonian, they induce error and noise to the hybrid optimization subroutine, making it harder and harder to find good quality solutions. And finally, because of this iteratively complicating concept, with a high change, the first feasible solution we find is the best that the algorithm can achieve.

\subsection{Algorithmic scheme}

\begin{figure}[h!]
    \centering
    \begin{tikzpicture}[
        node distance=1.2cm, 
        every node/.style={draw, rectangle, minimum width=9cm, minimum height=0.8cm, align=center},
        >=stealth 
    ]
        \node (top)[thin] {Construct initial RPH (\ref{sec:RPH})};
        \node[below of=top, thin] (second) {Solve the Ground State Problem on (RPH) (\ref{sec:qsolve})};
        \node[below of=second, thin] (third) {Sample from the result state (\ref{sec:sampling})};
        \node[below of=third, very thick] (fourth) {Evaluate samples. Check stopping criteria. (\ref{sec:end})};
        \node[below of=fourth, thin] (fifth) {Compute violation scores (\ref{sec:vscore})};
        \node[below of=fifth, thin] (sixth) {Add new constraints to (RPH) (\ref{sec:addcon})};
        
        \draw[->] (top.south) -- (second.north);
        \draw[->] (second.south) -- (third.north);
        \draw[->] (third.south) -- (fourth.north);
        \draw[->] (fourth.south) -- (fifth.north);
        \draw[->] (fifth.south) -- (sixth.north);
        
        \draw[->] (sixth.east) -- ++(0.7, 0) -- ++(0, 4.8)  -- (second.east);
    \end{tikzpicture}
    \caption{Steps of the constraint generation scheme}
    \label{fig:scheme}
\end{figure}
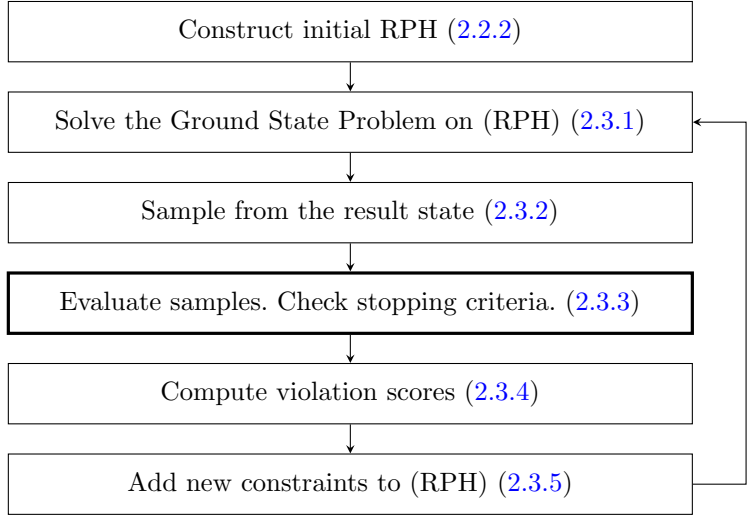

We define a constraint generation scheme, where we iteratively update the Hamiltonian (RPH) performing a run of a (hybrid) quantum optimization subroutine on it, as shown on Figure \ref{fig:scheme}. We first initialize our problem in (RPH) with $\hat{A}^{(0)}$ and $\hat{b}^{(0)}$, that are all zeros. Then, we iterate the followings. Solve the Ground State Problem of (RPH) with a suitable subroutine, resulting in a state, which we then sample from. After that, we evaluate the samples, looking for feasible solutions, and on success, report the best feasible solution and optionally stop the run. If there are no more constraints to add because either each of the samples is feasible or violates only constraint that have been already added, stop the run and return the best found solution. Then, we compute constraint violation scores from the samples. Those are dual-like values that store information about which constraints are violated more frequently in (BLP). In the last step inside an iteration, based on the violation scores, we choose new constraints to add to $\hat A$ and $\hat b$ and update (RPH) accordingly.

\subsubsection{Solution from the Relaxed Problem Hamiltonian}
\label{sec:qsolve}

In the first step of the algorithm, we solve the ground state problem of (RPH) with an applicable algorithm $\mathcal{A}$. Formally, $\mathcal{A}(H) = U|0\rangle^{\otimes n} = |\psi^*\rangle$ where an appropriate measurement on the computational basis would present a sample from the probability distribution presented by the solution state $|\psi^*\rangle$. We would like to highlight that the result of $\mathcal{A}$ is the state preparation routine $U$ which we can utilize to perform sampling from the state it prepares $|\psi^*\rangle$. 

\subsubsection{Sampling}
\label{sec:sampling}

 $\mathcal{A}$ can be obtained by, for example using a hybrid, parameterized quantum circuit based optimization algorithm, like the QAOA family, that obtains a final hyper-parameter set and a state preparation routine so we do not need to run the whole optimization process to reconstruct the final quantum state for each sample.

We obtain $q$ samples from the solution state. We denote the number of different samples by $s$ and their respective count in the samples by $\omega=(\omega_1,\ldots,\omega_s)$. 

To put the samples into our mathematical context, let us first denote all different samples by $X$, where sample $\ell \in \{1,\ldots,s\}$ is the $\ell$th column of $X$, i.e. where $X_{i\ell}$ is $x_i$ of sample $\ell$.

\subsubsection{Feasibility check}
\label{sec:end}

Then, we look for feasible solutions. Since with each constraint we add complicating terms to the Hamiltonian, the best feasible solutions from earlier steps of the constraint generation algorithm should correspond for better quality solutions to the (BLP). That might change as it depends on the quantum subroutine $\mathcal{A}$ and the noise characteristics of the physical hardware. From the mathematical algorithmic point this change is considered as error and can be computed separately for each algorithm $\mathcal{A}$ and hardware.

Based on this behavior, we define this step as a stopping criterion, by the following criteria as examples upon at least one feasible solution found: 

\begin{itemize}
    \item The solution value reaches a predefined threshold.
    \item The ratio of feasible samples over all samples reach a predefined threshold.
    \item All samples are feasible, thus no constraint to be added.
\end{itemize}

Formally, we gather the set of feasible sample indices as

\begin{equation}
    {S} = \left\{\ell\in\{1,\ldots,s\} ~\bigm|~ \sum_{i}A_{ji}X_{i\ell} = \mathbf{0} ~\right\}
\end{equation}
where $\mathbf{0}$ is the zero vector and $X_\ell, \ell \in {S}$ are the feasible solution vectors. Then, we select the best objective value as
\begin{equation}
    z = \min_{\ell \in {S}} c^TX_\ell,
\end{equation}
and set the best solution, $x^*=X_\ell$ where $z = c^TX_\ell$.

Since the above mentioned errors can occur, one might not stop the algorithm here, but continue and look for better solutions in latter iterations until no constraint can be added anymore. 

\subsubsection{Violation scores}
\label{sec:vscore}

First, we define the violation matrix $V \in \left\{0,1\right\}^{m \times s}$ 
\begin{equation}
    V_{j\ell} = \begin{cases}
        1, \text{when } \sum_i A_{ji}X_{i\ell} - b_j \neq 0 \\ 
        0, \text{otherwise.}
    \end{cases} 
\end{equation}
Then, counts of violations for each constraint are given as the row-sums of $V$. Since we aim for a relative quantity, we normalize it by the sample size. Therefore, the violation score vector $\nu \in \mathbb{R}^m, \nu_j \in \left[0, 1\right]$ is given by
\begin{equation}
    \nu_{j} = \frac1{q}\sum_\ell V_{j\ell}\omega_{\ell}. 
\end{equation}

\subsubsection{Constraint updates}
\label{sec:addcon}

One could think of the violation score vector $\nu$ as dual-like quantities: they measure the likelihood of the constraints to be violated upon sampling from the hybrid optimization subroutine's solution state.

Based on the violation scores, let us now determine which constraints, or in physical sense, coupling terms to add to (RPH). Our aim is to reduce the number of violations while keeping (RPH) as simple as possible.

We now define a threshold value $t \in \left[0, 1\right]$ for $\nu$. It determines how strongly violated constraints to add, which affects how many constraints we add. In its corner cases, $t=0$ means we add all constraints at once, while $t=\norm{\nu}_\infty$ we add only the constraint with the highest violation score. Note that this ensures a complete optimization routine only if the quantum hardware provides reasonable solutions. If the quantum subroutine fails to satisfy existing constraints, i.e. to find the low energy states, there might be no new constraint to add and thus the constraint generation could fail. This is an unfortunate case where the underlying quantum algorithm would also fail to provide reasonable results, so our strict improvement claim (\ref{sec:res-theo}) holds. However, from a practical standpoint, one could resolve it by artificially lowering the value of $t$ whenever no constraint could be added, nor feasible solution has been found. Since selecting the right $t$ value is important, we discuss it further in Appendix \ref{app:tvalue}.

For the $k$th iteration, we collect the new constraints that were not included yet to the (RPH) in
\begin{equation}
    R^{(k)} = \left\{ j \in \{1,\ldots,m\} \mid \hat A^{(k)}_j = 0\right\}
\end{equation}
and the constraint indices that we would like to add are gathered in a binary vector $\tau^{(k)} \in \{0,1\}^m$, where 
\begin{equation}
    \tau^{(k)}_j = \begin{cases}
        1 \text{ if } \nu_j^{(k)} \geq t \text{ and } j \in R^{(k)} \\
        0~ \text{otherwise.}
    \end{cases} 
\end{equation}
Then, our new penalty matrix $\hat A^{(k)}$ is given by
\begin{equation}
    \hat{A}^{(k)} = \hat{A}^{(k-1)} + diag(\tau^{(k)})A
\end{equation}
where $diag(\tau^{(k)})$ is the square diagonal matrix, created from $\tau^{(k)}$. Then, $\hat{b}^{(k)}$ is given by
\begin{equation}
    \hat{b}^{(k)} = \hat{b}^{(k-1)} + diag(\tau^{(k)}) b.
\end{equation}

\subsection{Implementation}

Using the steps from above as external functions, the algorithm is shown in Algorithm \ref{alg:congen}.

\begin{algorithm}
\caption{Constraint generation scheme}\label{alg:congen}
\begin{algorithmic}
\Require $c$, $A$, $b$
\State $\hat{A} \gets \mathbf 0_{n\times m}$
\State $\hat{b} \gets \mathbf 0_{m}$
\State $H$ $ \gets
\func{calculate\_hamiltonian}(c, \hat{A}, \hat{b})$ 
\Comment{see Section \ref{sec:RPH}}
\State $R \gets \{1,\ldots,m\}$
\While{ $|R|>0$}
    \State $|\psi^*\rangle \gets \mathcal{A}(H)$ 
    \State $X \gets \func{sample}(|\psi^*\rangle, s)$ \Comment{see Section \ref{sec:sampling}}
    \If{$x \gets \func{feasibility\_check}(X)$} \Comment{see Section \ref{sec:end}}
        \Return $x$
    \EndIf
    \State $\nu \gets \func{violation\_scores}(X, A, b)$  \Comment{see Section \ref{sec:vscore}}
    \State $\hat{A}, \hat{b} \gets \func{add\_constraints}(\hat{A}, \hat{b}, \nu, A)$ \Comment{see Section \ref{sec:addcon}}
    \State $ H  \gets \func{calculate\_hamiltonian}(c, \hat{A}, \hat{b})$ \Comment{see Section \ref{sec:RPH}}
\EndWhile
\end{algorithmic}
\end{algorithm}

There are many opportunities in the details to create an efficient implementation. One could utilize the sparsity of $\hat{A}$ and $\hat{b}$.

We implemented the algorithm in Python 3.13. We are aware of how weak the performance of Python is operationally, compared to compiled languages. However, with outsourcing potentially computation-heavy tasks to optimized, compiled packages greatly mitigates this issue. In fact, in this case, the quantum solver routine dominates all other steps.

\subsection{The quantum subroutine}

The quantum subroutine is an essential but hidden element of the constraint generation scheme. The practical results highly depend on it. For simplicity and mainly because of classical simulation resource constraints, we used the original QAOA \cite{Farhi2014} algorithm. Since it is a variational quantum algorithm (VQA) in which a problem with similar structure is solved at each step, we exploited the parameterization by explicitly setting initial hyperparameters (see Appendix \ref{app:initparams}) for the subroutine. While the QAOA algorithm has its own problems, it is a simple and suitable choice to demonstrate our algorithm with. In general, the better the quantum algorithm is, the better the results that we could achieve. This comes from our claim that up to the probabilistic behavior of the VQA, the constraint generation produces at least as good quality solutions as the quantum subroutine itself.

\section{Results}
\label{sec:res}

\subsection{Theoretical aspects}
\label{sec:res-theo}

Our theoretical result relies on the tradeoff between simplicity and feasibility. That is, on one hand, relaxed problems are easier to solve. The Hamiltonian operator is sparse and the quantum algorithm has a less complicated structure to explore. On the other hand the relaxed problem has a much larger feasible area. Even if the quantum algorithm finds good, or optimal solutions to it, they might be far from feasibility for the original BLP. 

Another consequence of this tradeoff is that, the quantum algorithm becomes less accurate with regard to optimality of the relaxed BLP, while more and more feasible solutions appear. This is due to the fact that as we progress, less and less constraints are relaxed.

Ideally, after some iterations and before the whole set of constraints added, we provide enough penalties to enforce the highest number of feasible solutions at the sampling step, while the Hamiltonian is still not too complicated to optimize over. Before that point, most found solutions are infeasible, and after it, the performance of the VQA drops as RPHs become more complex.

\begin{figure}[ht]
    \centering
    \includegraphics[width=1.0\textwidth]{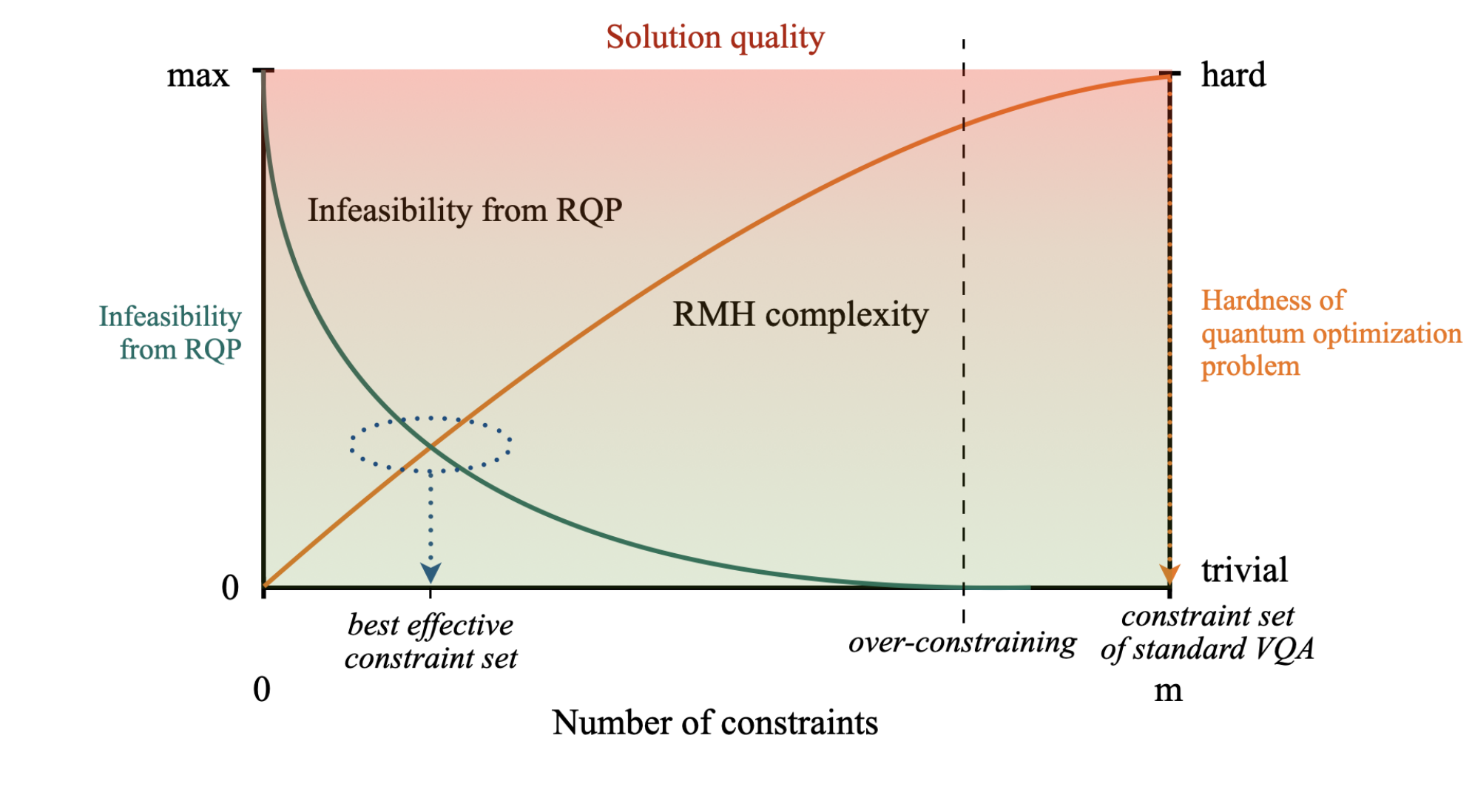}
    \caption{Illustrative figure of the tradeoff between the error coming from relaxing constraints and the error coming from the hardness of training VQAs. The gradient coloring corresponds for expected solution values for the BLP minimization problem.}
    \label{fig:congen_elm}
\end{figure}

In the reasoning above, we used two facts. First is a monotonic increase of RPH complexity over the iterations, and thus monotonic decrease in solution quality of the quantum algorithm. Second is a monotonic improvement in the feasibility of the solutions  of the RHPs, regarding to the BLP. The corresponding formal statements and  proofs can be found in Appendix \ref{app:proofs}.

In Figure \ref{fig:congen_elm} we show how finding the optimal set of constraints could provide better quality solutions than the original VQA on the complete problem. The more constraints the problems have relative to the number of variables, the smaller is the best effective constraint set compared to the number of all constraints. Visually, that brings the best effective set of constraints more to the left side.

The affect of the chosen value of $t$ can be interpreted via the figure. It is then a step size on the horizontal axis. Choosing a small step size ensures that we can find the best effective constraint set, however, with many iterations. On the other side, with choosing a large step size one reduces chance to find a small effective constraint set. We discuss it in Appendix \ref{app:tvalue}.

From our first statement, less constraints entail better quality solutions, which also implies that our constraint generation scheme produces at least as good quality solutions as the quantum algorithm it wraps, up to the probabilistic behavior of the VQA.

\subsection{Practical results}
\label{sec:pracres}

We tested on many different instances of the weighted exact set covering problem. The problem is given by a set of elements $\mathcal{U}$, subsets $\mathcal{S} \subseteq P(\mathcal{U})$ of its power set, $\mathcal{S}=\{S_1,\ldots,S_n\}$, and their corresponding weights $w_1, \dots, w_n$. Then, the problem is to select a set of subsets $Q$ from $\mathcal{S}$ that cover exactly $\mathcal{U}$, i.e. $\bigcup_{S_i \in Q} S_i = \mathcal{U}$ and $S_i \cap S_j = \emptyset$ for each $S_i, S_j \in Q, i\neq j$. The objective is to also minimize their total weight $\sum_{S_i\in Q} w_i$. 

As a binary linear program, for variables $x_i$ that encode decisions whether we select $S_i \in \mathcal{S}$ into our solution or not,
\begin{align}
    &\min \quad \sum_{S_i\in\mathcal{S}} w_i x_i \\[6pt]
    &~s.t. \quad \sum_{S_i\in\mathcal{S}\ :\ e \in S_i} x_i = 1 \quad &\forall e \in \mathcal{U} \\[6pt]
    &\qquad \quad x_i \in \{0,1\} &\forall S_i \in \mathcal{S}
\end{align}

We use the standard QAOA algorithm \cite{Farhi2014} within constraint generation and also for comparison. One might use more sophisticated algorithms \cite{Chalupnik2022, hegade2022, romero2024} and real quantum hardware to fully utilize the capabilities of our work. We used simulators through the Qiskit \cite{qiskit2024} Python package to simulate quantum subroutines.

Since quantum algorithms do not prove optimality in practice, we solved the problem instances with the open source solver SCIP \cite{BolusaniEtal2024OO} to obtain reference results. The QAOA we use for comparison and for constraint generation have the same structure, with $p=2$, and run on the same simulated hardware. The QAOA itself can be interpreted as the $t=0$ version of the constraint generation scheme, where we immediately add all the constraints after a first sampling iteration.

We tested the algorithm with random problems of 4 different configurations sets. The evaluations are aggregated comparisons over 100 problems in each case. The quality is determined by calculating how close is the quantum algorithm to the optimal solution value, converted to percentage. That is, $\frac{\text{opt}}{\text{alg}} \times 100$ as it is a minimization problem. Table \ref{table:avgval} shows how close the solution values were to the optimal. In Table \ref{table:feassol}, we show how many problems the respective algorithms could solve. Detailed descriptions of the problems and results are provided in Appendix \ref{app:res}. 

\begin{table}[h]
\centering
\caption{Average objective value in percentage of the optimal value. Calculation is based on approximation ratio and averaged over the problems in each problem set.}
\label{table:avgval}
\begin{tabular}{lcccc}
\hline
Method & WEC-8-25-12 & WEC-10-20-18 & WEC-12-40-14 & WEC-12-40-15 \\
\hline
QAOA & 70.97\% & 25.86\% & 10.67\%  & 30.66\% \\
Constraint Generation & 84.97\% & 57.85\% & 58.25\%  & 77.29\% \\
\hline
\end{tabular}
\end{table}

\begin{table}[h]
\centering
\caption{Percentage of feasible solutions found over problems in each problem set.}
\label{table:feassol}
\begin{tabular}{lcccc}
\hline
Method & WEC-8-25-12 & WEC-10-20-18 & WEC-12-40-14 & WEC-12-40-15 \\
\hline
QAOA & 75\% & 36\% & 12\% & 38\% \\
Constraint Generation & 99\% & 84\% & 70\% & 98\% \\
\hline
\end{tabular}
\end{table}

\section{Discussion}
\label{sec:discussion}

The overall performance of the constraint generation highly depends on the quantum algorithm it encapsulates. We used QAOA inside constraint generation and for comparison to experimentally show the improvement. While comparison is fair and supports our claim, it is worth to note that it is only a proof of concept. Since any improvement in the quantum algorithm would directly improve the constraint generation, selecting the right subroutine and optimizing it for the purpose is essential.

This algorithm is not a new idea by any means. However, it uses ideas from fields that are established, but still far from each other. There is a conceptually similar algorithm family introduced recently, named QIRO \cite{Fingar2024}, that also works by wrapping quantum optimization algorithms into classical decomposition framework. However, there are many differences: QIRO is a top-down algorithm family, starting with the complete problem and reducing it at each step. Our constraint generation framework is a bottom-up concept, starting with the simplest problem and then iteratively adding complexity to it. Furthermore, our algorithm is theoretically an exact optimization algorithm for binary linear problems. It can be used as a primal heuristic, or as complete optimization algorithm for certain problems.

Constraint generation is a well-known concept in OR. Usually when the number of constraints is large, dropping a well defined set of constraints  simplifies the problem while still maintaining its initial structure. Also, many algorithms that are in the toolbox of an OR practitioner, are exact and have great convergence analysis that might be a great proven enhancement for quantum algorithms. That is why we indicate that our aim with this work is to show the conceptual approach, and the constraint generation itself is a first step and good example of showing its potential.

Setting the right $t$ value is important for practical reasons: solving simpler problems provide better quality results from the quantum algorithm. From a simulation perspective, it also reduces resource requirements significantly. As a future work, another interesting direction could be to set $t=\varepsilon, 1/q \geq \varepsilon > 0$, where  $\varepsilon$ is small enough, which would imply adding all violated constraint at each step. That is a common practice in constraint generation literature in OR.

The concept of bringing such fields together is also not new: OR and Machine Learning (ML) were initially distant, with their communities being closed toward each other. Then, it has changed, which provided a huge boost to optimization solvers. \cite{Cappart2023, Gasse2019, Gupta2020, FischettiLodi2011}
Since the Quantum Optimization is already advancing quite fast, deep collaboration between the respective fields is essential to take one more step forward. There are also some noteworthy advances looking into this direction in \cite{Naghmouchi2024, Ajagekar2022, Tang2024}.

\section{Conclusion}
\label{sec:conclusion}

The results show that our algorithm is more likely to find feasible solutions, and their value is on average better than the one that the reference finds. We emphasize that this reference could be any quantum algorithm, which produces a state preparation routine for its result state, that our algorithm encapsulates.

The results also show that the results of the reference and the constraint generation are close, when both of them find feasible solutions. Since the optimization algorithm is the same, this is expected, however the constraint generation algorithm has a slight advantage since the problem it solves is usually easier and less affected by noise. Since this is a weak simulated noise, the difference could be even clearer on larger instances and real quantum hardware.

\backmatter

\section*{Declarations}

\begin{itemize}
\item Funding: -
\item Conflict of interest/Competing interests: The authors declare that they have no competing interests.
\item Data availability: The implementation project with examples are available at https://github.com/tg90000/qcongen . 
\item Author contribution: A.C. created the mathematical framework, the implementation and testing, and wrote the main manuscript. B.G.-T provided insights to derivations and formal definitions throughout. All authors read and approved the final manuscript.
\end{itemize}

\bibliography{qcongen-bib}

\begin{appendices}

\section{Quadratic form to Hamiltonian}
\label{app:QUBOtoHAM}

The derivations of the Relaxed Problem Hamiltonian in Section \ref{sec:RPH} are the following.

Since the forms are already similar, we can do the following mapping of binary variables $x$ to new variables $\hat x \in \{1, -1\}^n$ by setting
$x = \frac{\hat x + \1}2$. Thus, 
%
\begin{align}
    H({x}) =~ & x^T(M\hat{A}^T\hat{A})x + (c^T - 2M\hat{b}^T\hat{A})x + M\hat{b}^T\hat{b} \\
    =~ & \frac{\hat x^T + \1^T}2(M\hat{A}^T\hat{A})\frac{\hat x + \1}2 + (c^T - 2M\hat{b}^T\hat{A})\frac{\hat x + \1}2 + M\hat{b}^T\hat{b} \\
    =~ & \frac{1}{4} M \hat{x}^T (\hat{A}^T \hat{A}) \hat{x} 
    + \frac{1}{2} M \mathbf{1}^T (\hat{A}^T \hat{A}) \hat{x} 
    + \frac{1}{4} M \mathbf{1}^T (\hat{A}^T \hat{A}) \mathbf{1} \\
    & ~+ \frac{1}{2}  c^T\hat{x} + \frac{1}{2} \mathbf{1}^T c
    - M  \hat{b}^T \hat{A} \hat{x} - M \hat{b}^T \hat{A} \mathbf{1}
    + M \hat{b}^T \hat{b} \\
    =~ & \frac{1}{4} M \hat{x}^T (\hat{A}^T \hat{A}) \hat{x} + \frac{1}{2} \left(c^T - 2 M\hat{b}^T \hat{A} + M\mathbf{1}^T (\hat{A}^T \hat{A}) \right) \hat{x} \\
    & ~+ \frac{1}{4} M \mathbf{1}^T (\hat{A}^T \hat{A}) \mathbf{1} + \frac{1}{2} \mathbf{1}^T c + M \hat{b}^T \hat{A} \mathbf{1}
    + M \hat{b}^T \hat{b}.
\end{align}
\noindent
Let us denote
\begin{equation}
    const = \frac{1}{4} M \mathbf{1}^T (\hat{A}^T \hat{A}) \mathbf{1} + \frac{1}{2} \mathbf{1}^T c + M \hat{b}^T \hat{A} \mathbf{1}
    + M \hat{b}^T \hat{b},
\end{equation}
which enables us to transform our problem to the form
\begin{equation}
    \min \left\{\frac{1}{4} M \hat{x}^T (\hat{A}^T \hat{A}) \hat{x} + \frac{1}{2} \left(c^T - 2 M\hat{b}^T \hat{A} + M\mathbf{1}^T (\hat{A}^T \hat{A}) \right) \hat{x} + const ~\bigm|~ \hat{x} \in \{-1,1\}^n \right\}.
\end{equation}
Then, for the spins $\sigma=(\sigma_1,\ldots,\sigma_n) \in \{-1,1\}^n$, the Ising Hamiltonian is given by
%
\begin{equation}
    H(\sigma) = - \sum_{ij}J_{ij}\sigma_i \sigma_j - \mu \sum_j h_i \sigma_i.
\end{equation}
So, we replace our auxiliary variables $\hat x_i$ with $\sigma_i, i=1,\ldots,n$, and the coefficients become in matrix form
\begin{equation}
    J = -\frac{1}{4}M\hat{A}^T\hat{A}
\end{equation}
and
\begin{equation}
    h = c^T - 2 M\hat{b}^T \hat{A} + M\mathbf{1}^T (\hat{A}^T \hat{A})
\end{equation}
with
\begin{equation}
    \mu = -\frac{1}{2}.
\end{equation}

Note that the Hamiltonian does not include the constant term. We can indeed put it aside for the quantum optimization subroutine, and add it back afterwards, as it does not affect the optimization process.

\section{Big M value}
\label{app:M}

The big $M$ value should be large enough to separate the set of feasible solutions from the set of infeasible solutions by penalizing the constraint violations. In general, it should be true that for any feasible $x$ and any infeasible $x'$, 
\begin{equation}
    c^Tx < c^Tx' + M\left(Ax'-b\right)^T\left(Ax'-b\right)
\end{equation}
which gives
\begin{equation}
    \frac{c^T\left(x-x'\right)}{\left(Ax'-b\right)^T\left(Ax'-b\right)} < M
\end{equation}
where $\left(Ax'-b\right)^T\left(Ax'-b\right) > 0$, as $x'$ is infeasible and it is like the squared distance from feasibility. 

It is important to find sufficiently large $M$, which is not larger than necessary to avoid numerical and physical issues. Formally it means,

\begin{equation}
    M > \max_{x, x'} \frac{c^T\left(x-x'\right)}{\left(Ax'-b\right)^T\left(Ax'-b\right)}
\end{equation}

One way to approach it would be to trying to minimize the denominator, while maximizing the nominator, keeping in mind that $\left(Ax=b\right)$ and  $\left(Ax'\neq b\right)$, $\forall x, x'$. That is,

\begin{equation}
\label{eq:app_M}
     \frac{\max_{x, x'} c^T\left(x-x'\right)}{\min_{x'}\left(Ax'-b\right)^T\left(Ax'-b\right)} \geq \max_{x, x'} \frac{c^T\left(x-x'\right)}{\left(Ax'-b\right)^T\left(Ax'-b\right)}.
\end{equation}
We now look for an upper bound for this expression. For the denominator of \eqref{eq:app_M}, we can in general avoid solving the computationally hard problem
\begin{equation}
    \kappa \leq \min_{x'} \left(Ax'-b\right)^T\left(Ax'-b\right),
\end{equation}
because $x'$ is binary, and $\kappa$ can be the numerical precision required to represent any element of $A$ or $b$. In general it is advisable to set it as large as possible. If, for example, the elements of $A$ and $b$ are integer, $\kappa$ can get the value 1. 

For the nominator of \eqref{eq:app_M} and once again using that $x, x'$ are binary
\begin{equation}
    c^T(x-x') ~=~ \sum_i c_i(x_i - x'_i) ~\leq~ \sum_i |c_i|. 
\end{equation}
When we combine the two, we get the following value for $M$:
\begin{equation}
    M ~=~  \frac{1}{\kappa}\sum_{i} |c_i| ~\geq~ \max_{x, x'} \frac{c^T\left(x-x'\right)}{\left(Ax'-b\right)^T\left(Ax'-b\right)}.
\end{equation}

\section{On the choice of $t$ value}
\label{app:tvalue}
The value of $t$ is important for the efficiency of the constraint generation. Here, we show how some corner values of $t$ makes the algorithm behave.

The first is, when $t=0$. In this case, all constraints added after the first, trivial iteration. This case is very close to just run the wrapped VQA straight on the problem. 

The second, when $t$ has a small, positive value, i.e. $t=\varepsilon, ~1/q \geq \varepsilon > 0$. Then, in an iteration, all constraints are added that have been violated by at least one sample. This is how a classical constraint generation algorithm would work --- by adding all violated constraints that had been previously relaxed. In our context, this is a greedy approach, where noisy samples could cause major issues. In this scenario, the samples are not aggregated, just checked which constraints they violate. That means an individual sample has greater responsibility, which requires reliability in the sampling process.

The third case is when $t$ is set to its maximum, i.e. $t=\norm{\nu}_\infty$, we add only one constraint, which had been violated most frequently by the samples. This is the single constraint that changes the most on the Hamiltonian with regard to the VQA final state measurement distribution. This is the constraint that drives the constraint generation most. However it is still only one constraint. There might be many other frequently violated constraints, and the algorithm might need many more iterations to finally produce feasible samples. It is a tradeoff between speed and accuracy.

Since the problem instances we can solve are very small, these characteristics are very hard to demonstrate, and thus this section is rather for discussion purposes.

\section{Formal support of the theoretical claims.}
\label{app:proofs}

In our theoretical results section we used two facts that we prove below.

\begin{statement}
    Let $\mathcal{A}$ be a VQA, and let RPH$^{(i)}$ and RPH$^{(i+1)}$ the Ising Hamiltonians of two successive iterations of the constraint generation algorithm. Then, 
    
    $$\frac{\mathcal{A}\left(\text{RPH}^{(i)}\right)}{OPT\left(\text{RPH}^{(i)}\right)} \leq \frac{\mathcal{A}\left(\text{RPH}^{(i+1)}\right)}{OPT\left(\text{RPH}^{(i+1)}\right)},$$
    where $\mathcal{A}(H)$  is the least energy value, obtained by a constant number of measurement on the final circuit of $\mathcal{A}$ and $OPT(H)$ is the ground state energy of $H$.
\end{statement}

\begin{proof}
    First, let us show that the number of many-body terms in RPH$^{(i)}$ is at most the number of many-body terms in RPH$^{(i+1)}$. The number of many-body terms in iteration $i$ is given by the number of non-zero elements above the main diagonal of $J^{(i)} = \left(\hat{A}^{(i)}\right)^T\left(\hat{A}^{(i)}\right)$. Let us suppose that $\hat{A}^{(i)}$ have $k$ non-zero rows (i.e. we already added $k$ constraints to the problem). Then, the number of non-zero elements of $J^{(i)}$, above its main diagonal is $NZ(J^{(i)})=k(k-1)/2 - \delta_k$, where $\delta_k$ is the number of orthogonal row-pairs in $\hat{A}^{(i)}$. That is,  $\boldsymbol{a_r}^T\boldsymbol{a_s} = 0$ for rows vectors $\boldsymbol{a_r}, \boldsymbol{a_s}$ of $\hat{A}^{(i)}$. Then, in iteration $i+1$ we add $l$ more rows, obtaining $\hat{A}^{(i+1)}$. The number of non-zeros above the main diagonal of $J^{(i+1)}$ is then $NZ(J^{(i+1)})=(k+l)((k+l)-1)/2 - \delta_k - \delta_l - \Delta_{kl}$, where $\Delta_{kl}$ are the number of orthogonal row pairs, where exactly one of the rows is from the first $k$ constraints, and one is from the newly added $l$.
    The difference between non-zeros of $J^{(i)}$ and $J^{(i+1)}$ is $kl + l(l-1)/2 - \delta_l - \Delta_{kl}$. The maximal value of $\delta_l$ is the number of all pairs from the new constraints  $\delta_l \leq l(l-1)/2$. Similarly, the maximal value of $\Delta_{kl}$ is the number of pairs, where there is exactly one row from the first $k$ nonzero rows, which gives $\Delta_{kl} \leq kl$. Thus,
    \[
NZ(J^{(i+1)}) - NZ(J^{(i)}) = kl + l(l-1)/2 - \delta_l - \Delta_{kl} \geq 0
    \]
This is difference in the number of many-body terms in RPH$^{(i)}$ and RPH$^{(i+1)}$. 

    Let us now take a step aside. On certain graph problems, density of the graph directly corresponds for the hardness of training VQAs, shown experimentally in \cite{Akshay2021}. The graph density on these problems directly correspond for the number of many-body terms in the Hamiltonian they encode the problem into. That is, an Ising Hamiltonian with more many-body terms is harder to optimize on with VQAs.

    This means that $\mathcal{A}(H)$ produces better results on Ising Hamiltonians with less many-body terms. Since \cite{Akshay2021} is an experimental result, where a constant number of measurements applied to the result state of the VQA, our statement can not be stronger than that. A better result in our context is being closer to the optimum, i.e. $\frac{\mathcal{A}\left(\text{H}\right)}{OPT\left(\text{H}\right)}$ is closer to 1. 

    Since RPH$^{(i)}$ has at most as many many-body terms as RPH$^{(i+1)}$, \cite{Akshay2021} shows evidence for 
    $$\frac{\mathcal{A}\left(\text{RPH}^{(i)}\right)}{OPT\left(\text{RPH}^{(i)}\right)} \leq \frac{\mathcal{A}\left(\text{RPH}^{(i+1)}\right)}{OPT\left(\text{RPH}^{(i+1)}\right)}.$$

\end{proof}

\begin{statement}Assume that a VQA $\mathcal{A}$ can find the feasible region of the given RPH it solves, formally, for the corresponding RQP and samples $X$ of $\mathcal{A}$(RPH) it holds that RQP($X_\ell$) $< M ~\forall \ell$. Furthermore assume that $X$ contains samples, uniformly distributed, from the feasible region of RQP. Let $\pi(X)$ denote the expected value of the number of the samples in $X$ that are feasible for the original BLP. Then, for two successive iterations $i$ and $i+1$, $\pi(X^{(i)}) \leq \pi(X^{(i+1)})$.
\end{statement}

\begin{proof}
    Informally, the statement says, in later iterations with more constraints, we are more likely to measure solutions that are feasible for the original BLP.
    To show that, we need to show that the feasible region of RQP$^{(i+1)}$ monotonically converges to the feasible region of BLP. For the expected value calculation, let $\mathcal{S}_{\text{BLP}}$ denote the set of  feasible solutions of BLP and $\mathcal{S}_{\text{RQP}^{(i)}}$ denote the set of solutions $x$ of RQP$^{(i)}$ where RQP$(x) < M$. We assumed that the samples are from a uniform distribution of $\mathcal{S}_{\text{RQP}^{(i)}}$, therefore the probability of a single sample being feasible for both RQP$^{(i)}$ and BLP is $|\mathcal{S}_{\text{BLP}}| / |\mathcal{S}_{\text{RQP}^{(i)}}|$. Since we do not assume the samples to be different, the expected value of feasible samples over $q$ measurements is  
    \begin{equation}
    \label{eq:piXi}
        \pi(X^{(i)}) = q \frac{|\mathcal{S}_{\text{BLP}}|}{|\mathcal{S}_{\text{RQP}^{(i)}}|}.
    \end{equation}
    To see how $\pi(X)$ changes over the iterations, we need to look at the only changing term above, $\mathcal{S}_{\text{RQP}^{(i)}}$. This is the set of feasible solutions for a relaxed BLP from iteration $i$. To get RQP$^{(i+1)}$, we add constraints. These constraints either cut the feasible region further, or they do not cut. Due to the impreciseness of VQAs, we do not guarantee that we only add constrants that are all restrictive on RQP$^{(i)}$, nor RQP$^{(i+1)}$. However we do guarantee that they do not allow more feasible solutions. Formally, $\mathcal{S}_{\text{RQP}^{(i+1)}} \subseteq \mathcal{S}_{\text{RQP}^{(i)}}$. Hence, $|\mathcal{S}_{\text{RQP}^{(i+1)}}| \leq |\mathcal{S}_{\text{RQP}^{(i)}}|$ follows. And if we combine it with \eqref{eq:piXi}, we arrive at
    \begin{equation}
        \pi(X^{(i)}) = q \frac{|\mathcal{S}_{\text{BLP}}|}{|\mathcal{S}_{\text{RQP}^{(i)}}|} \leq 
        \frac{|\mathcal{S}_{\text{BLP}}|}{|\mathcal{S}_{\text{RQP}^{(i+1)}}|} = \pi(X^{(i+1)}),
    \end{equation}
    which completes the proof.
\end{proof}

Note that our assumptions are strong and practically unrealistic. However, we aim to theoretically show how the iterations drive the reduced constraint set towards feasibility. If the VQA successfully finds the low-energy region of RPH, which corresponds to the feasible solutions, but unable to explore the smaller differences between its energy levels, this becomes a reasonable approximation. The issue from this approximation is, we do not consider the re-addition or weighing of constraints that have been already added, but still not satisfied. That can be a path of a future work.

\section{Variational initial parameters for iterations}
\label{app:initparams}

The initial parameters for the quantum subroutine are important. We solve the original (BLP) iteratively, thus keeping parameters from previous constraint generation iterations might provide helpful information.

We did not go deep into this direction of hyperparameter optimization of the different VQAs. However we did keep the optimal parameter set from the last iteration and used as initial parameters in the current iteration. 

\section{Experimental results}
\label{app:res}

Our implementation uses simulated quantum hardware through the Qiskit SDK from IBM.

The problems range from 8 variables to 12 variables with 20 to 40 constraints. The problems are generated randomly with integer $w_i \in \text{[}1,100\text{]}$ and variable set sizes. For each test set, the number of variables (subsets) and constraints (elements) are fixed. The achieved values are relative gaps to the optimal solution, i.e.\ $\frac{\text{opt}}{\text{alg}} \times 100$. We note that our plots do not compare results directly on each instance, and they show only aggregations. Note that on some instances, the QAOA might have performed better, due to its probabilistic nature, and to some extent due to errors coming from the computation itself. In the results below, we aim to present the overall quality of the solutions found, showing that on average, our approach is better. 

\subsection{Tiny problems: WEC-8-25-12}

\begin{figure}[ht]
    \centering
    \includegraphics[width=0.9\textwidth]{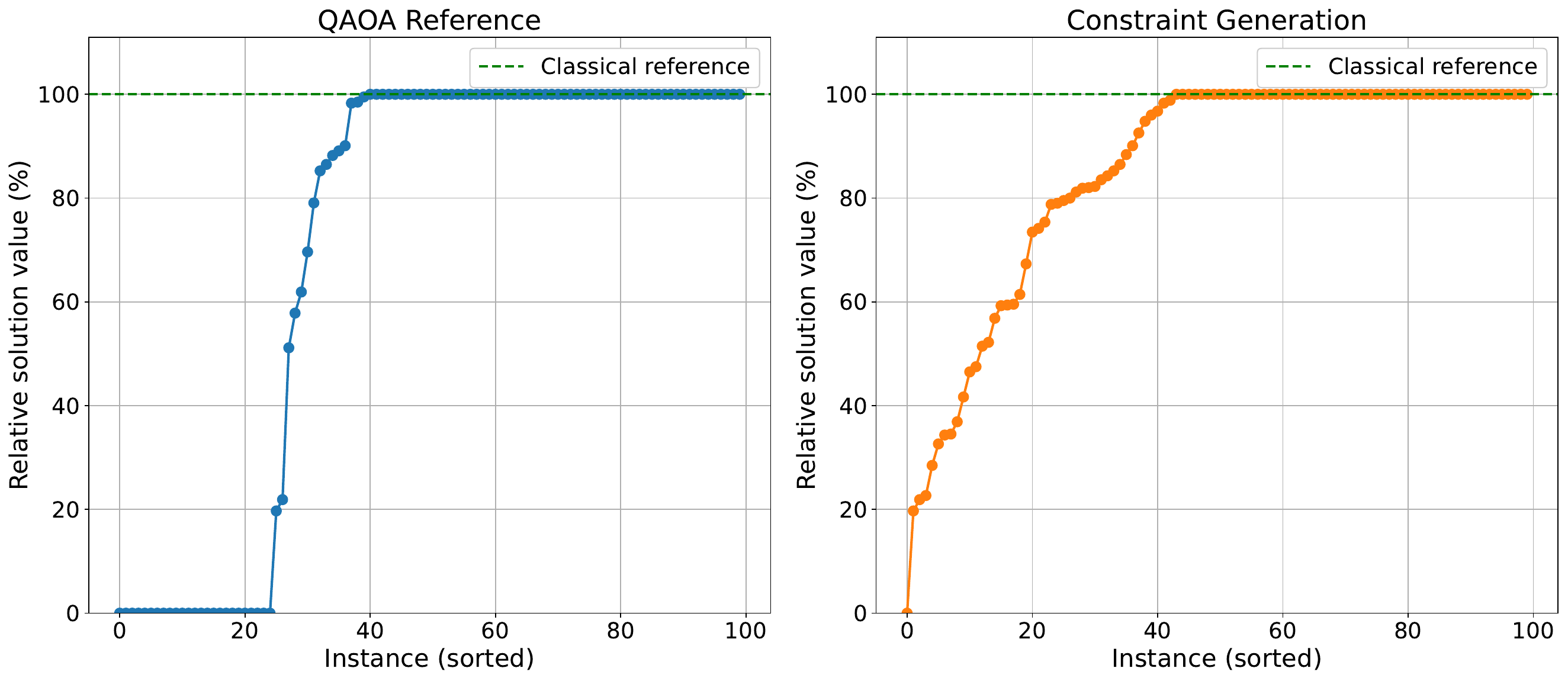}
    \caption{Gaps to optimal solutions by problem, sorted by value, on EC-8-25-12}
    \label{fig:sorted825}
\end{figure}

The first problem set had 8 subsets, 25 elements in the universe. The sets in the problems had cardinality between 1 and 12, creating a diverse set of problems. We call them WEC-8-25-12. In general, a slight advantage can be seen as our algorithm finds on average better solutions more frequently, see Figure \ref{fig:sorted825}.

\subsection{Small problems: WEC-10-20-18}

\begin{figure}[ht]
    \centering
    \includegraphics[width=0.9\textwidth]{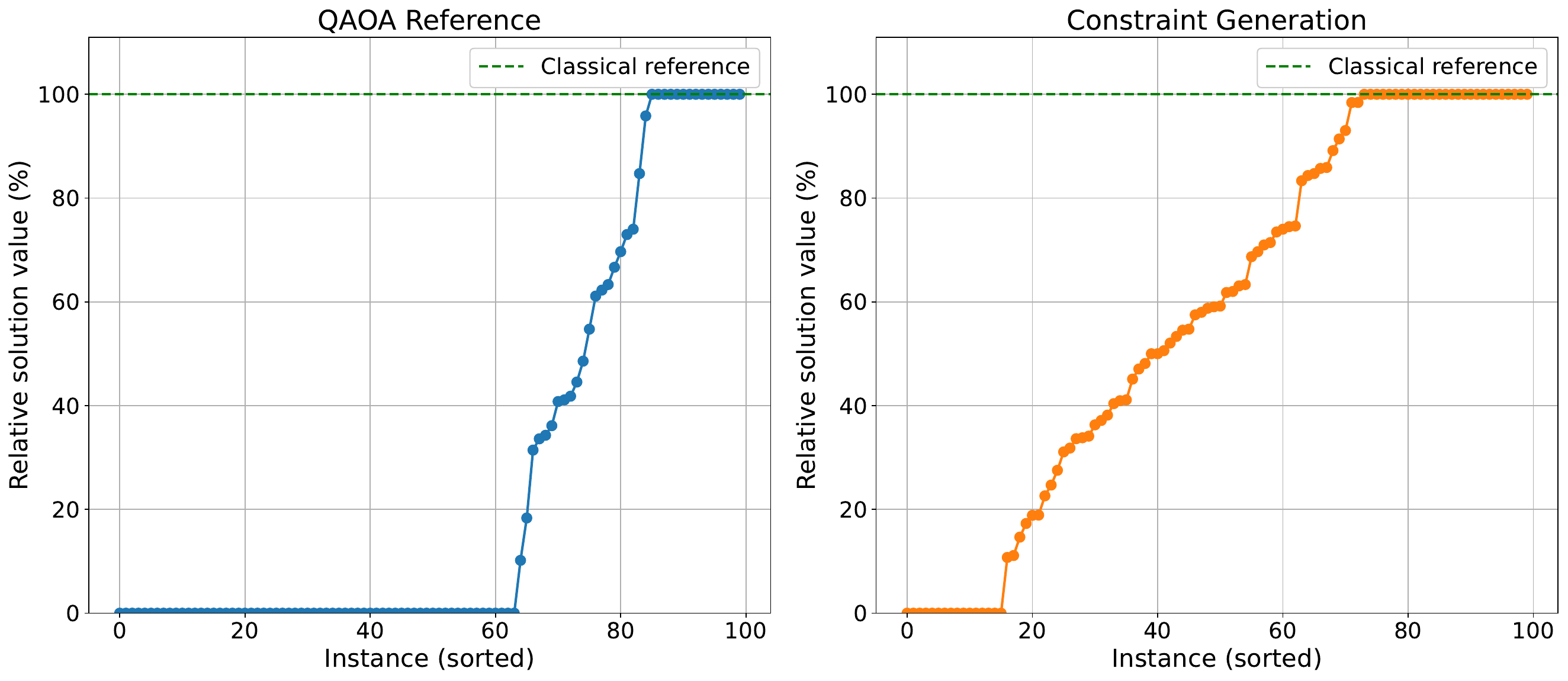}
    \caption{Gaps to optimal solutions by problem, sorted by value, on WEC-10-20-18.}
    \label{fig:sorted1020}
\end{figure}

The second problem set had 10 subsets and 20 elements in the universe. The sets could at most have 18 elements, hence the name WEC-10-20-18. The comparison shows that our algorithm finds feasible solutions for most problems, unlike QAOA. Therefore the overall solution quality is higher as well, which can be observed on Figure \ref{fig:sorted1020}.

\subsection{More variables: WEC-12-40-14, WEC-12-40-15}

\begin{figure}[ht]
    \centering
    \includegraphics[width=0.9\textwidth]{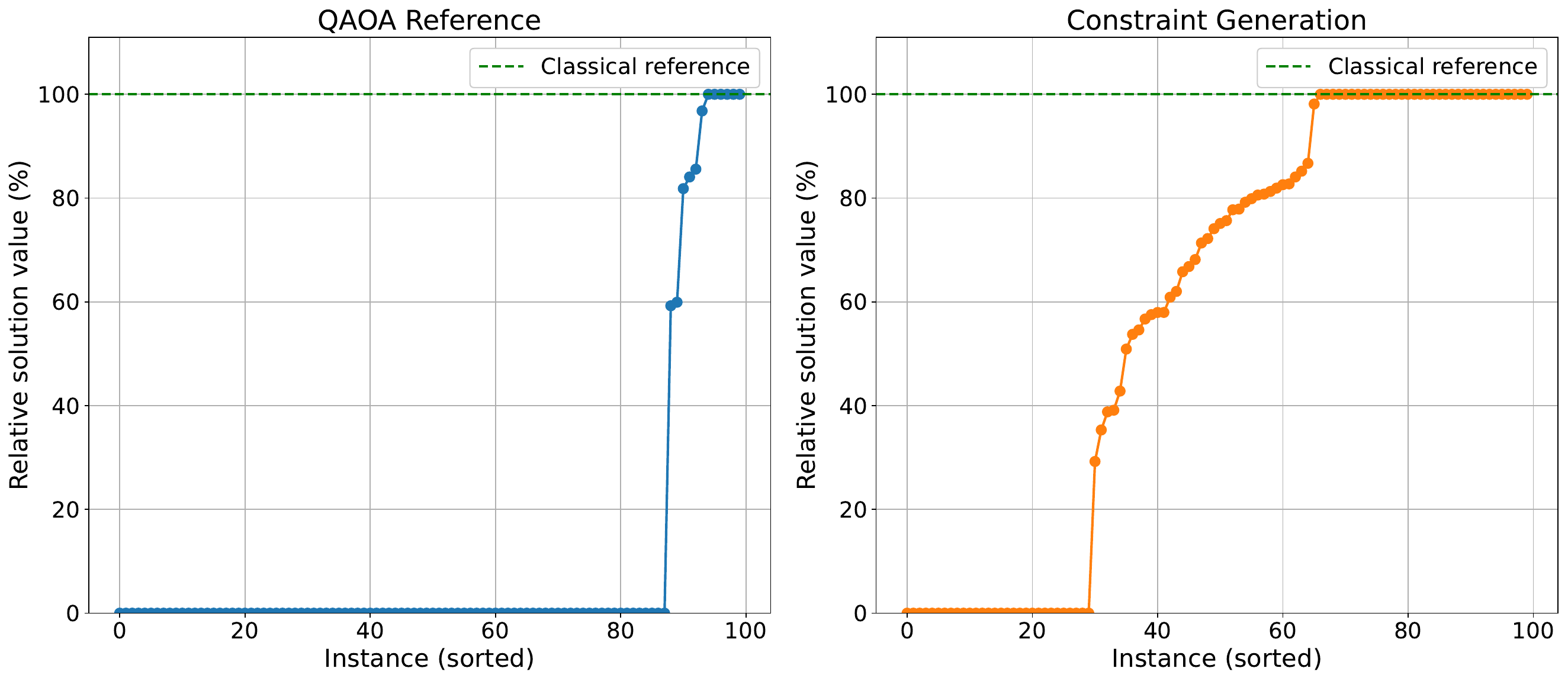}
    \caption{Gaps to optimal solutions by problem, sorted by value, on WEC-12-40-14.}
    \label{fig:sorted124014}
\end{figure}

\begin{figure}[ht]
    \centering
    \includegraphics[width=0.9\textwidth]{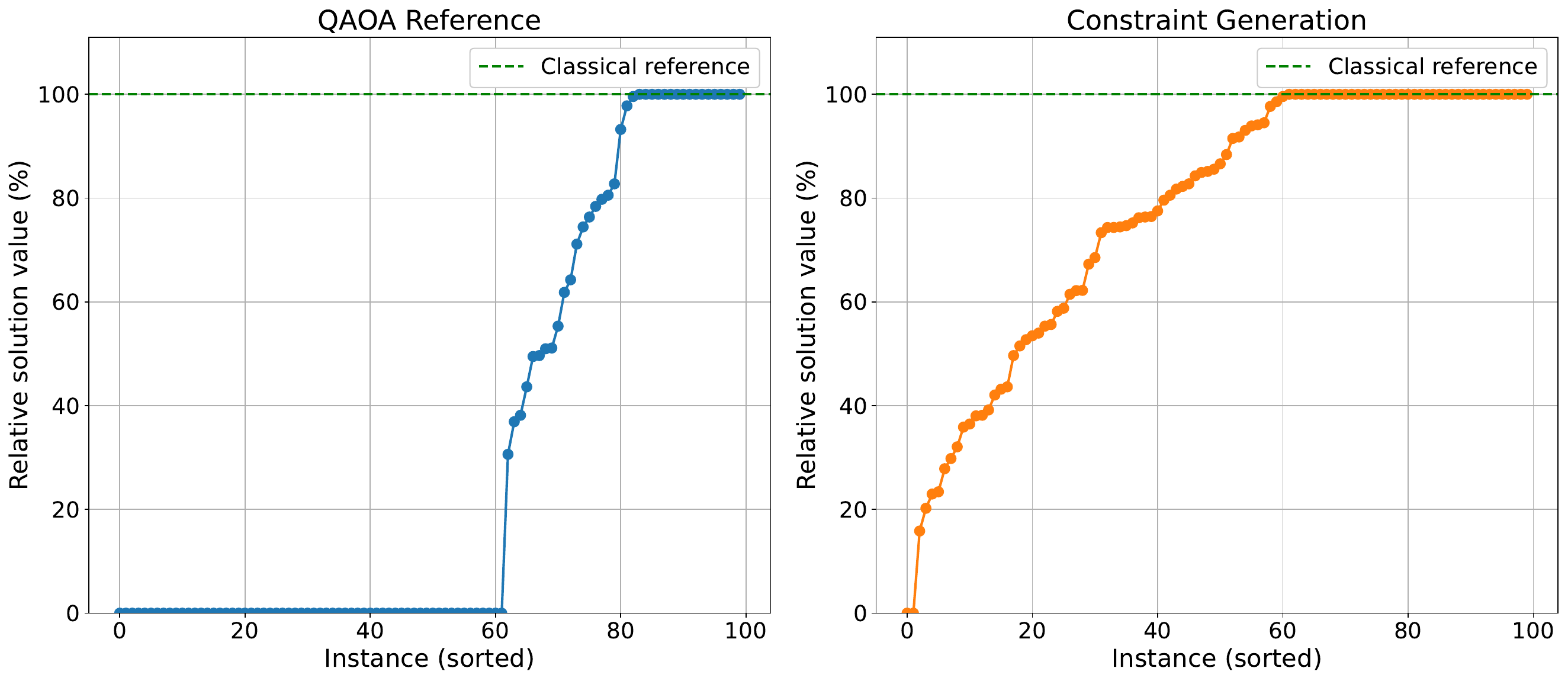}
    \caption{Gaps to optimal solutions by problem, sorted by value, on WEC-12-40-15.}
    \label{fig:sorted124015}
\end{figure}

The third and fourth problem sets had 12 subsets, and 40 elements in their universe. Sets could have at most 14 elements for problems WEC-12-40-14 and 15 elements for problems WEC-12-40-15. The gap between QAOA and our algorithm is even larger, mostly in terms of feasible solutions found. These results are presented in Figures \ref{fig:sorted124014} and \ref{fig:sorted124015} respectively.

\subsection{Different number of constraints}

It has been observed that more dense Hamiltonians provide less accurate results with VQAs \cite{Akshay2021}. Since we iteratively add the constraints, the density of the relaxed problem grows by each iteration, we expect that our constraint generation scheme has a greater advantage on BLPs with more constraints.

Our numerical experiments support this. Here, we created 30 problem instances with 8 variables, for different number of constraints. Then, to get one data point, we took the average of the instances for one algorithm, for one constraint set size. This is necessary to show interpretable results, because of the probabilistic nature of VQAs. 

\begin{figure}[ht]
    \centering
    \includegraphics[width=0.9\textwidth]{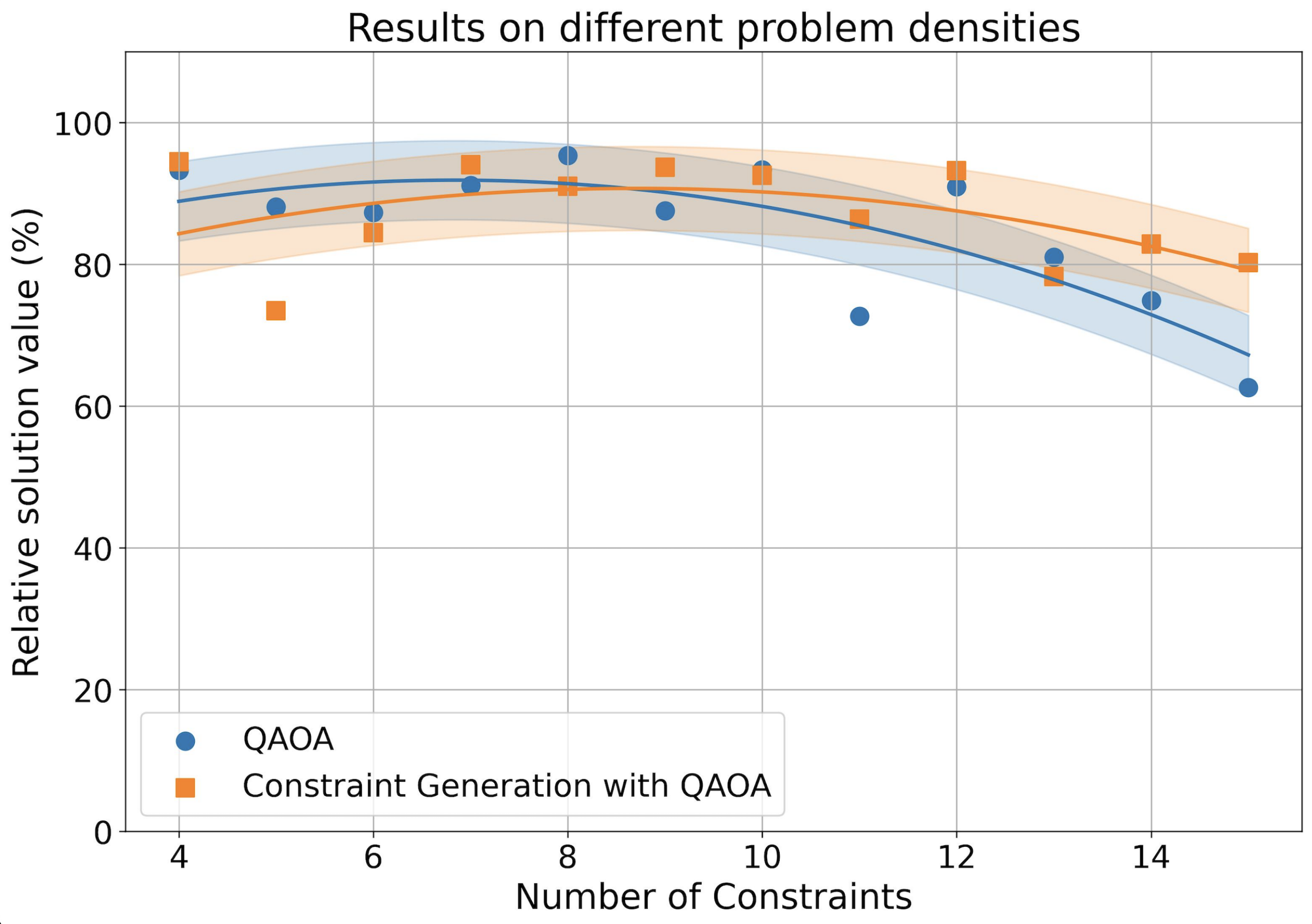}
    \caption{The relative solution value on problem with different number of constraints.}
    \label{fig:densities}
\end{figure}

In Figure \ref{fig:densities} we show the average results for each problem set with orange and blue dots. The curves are fitted to the respective points. The colored area next to the curves are the standard deviation of the averages from the curve. Even though the example is tiny, the growth of the difference, as the problems include more and more constrains, is observable.

\end{appendices}

\end{document}